\documentclass[11pt,twoside]{article}
\usepackage{asp2004}
\usepackage{psfig}
\usepackage{epsf}
\usepackage{graphics}
\usepackage{lscape}
\def\edcomment#1{\iffalse\marginpar{\raggedright\sl#1\/}\else\relax\fi}
\reversemarginpar

\begin{document}
\title{Dynamic Tides and the Evolution of Stars in Close Binaries}
\author{B. Willems$^1$ and A. Claret$^2$}
\affil{$^1$Northwestern University, Department of Physics and Astronomy,
  2145 Sheridan Road, Evanston, IL 60208, USA}
\affil{$^2$Instituto de Astrof{\'\i}sica de Andaluc{\'\i}a, CSIC, Apartado
  3004, 18080 Granada, Spain}

\begin{abstract}
In this talk, we review some recent advances in the theory of dynamic
tides in close binaries. We particularly focus on the effects of
resonances of dynamic tides with free oscillation modes and on the
role of dynamic tides in the comparison of theoretically predicted
and observationally inferred apsidal-motion rates. 
\end{abstract}
\thispagestyle{plain}

\section{Introduction}

In close binaries, the tidal force exerted by each star on its
companion perturbs the shape of the stars from spherical
symmetry. Depending on whether or not the binary orbit is circular and
the component stars are rotating synchronously with the orbital
motion, the perturbations will be static (equilibrium tides) or time
dependent (dynamic tides). In the latter case, the tidal force induces
forced nonradial oscillations in the component stars, which may be
resonant with the components' free modes of oscillation (Cowling
1941).

The tidal deformation of a binary component furthermore perturbs the
spherical symmetry of the star's external gravitational field. For
eccentric binaries, this perturbation gives rise to a secular shift of
the line of the apses, the period of which depends on the orbital
parameters and on the internal structure of the component
stars. Comparison of apsidal-motion rates derived from theoretical
stellar models with observationally inferred apsidal-motion rates
therefore provides an excellent tool to test current theories of
stellar structure and evolution. Until recently, the comparison was
usually carried out in the static-tide approximation to avoid the
complications of a time-dependent tidal potential.

Claret \& Willems (2002) were the first to include the effects of
dynamic tides in a comparison between theoretically predicted and
observationally derived apsidal-motion rates for a large and
homogeneous sample of eccentric eclipsing binaries. The authors found
that when the sample is restricted to systems with accurately known
stellar and orbital dimensions, the comparison between theory and
observations showed an excellent degree of agreement. However, despite
this general agreement, some systems (e.g. DI\,Her, AS\,Cam,
V451\,Cyg) still show apparently anomalous apsidal-motion rates which
are not fully understood yet. Continuing efforts in improving both
theory and observations are therefore essential to further increase
the merit of the apsidal-motion test to stellar structure and
evolution.

In the following sections we outline some basic aspects of the theory
of dynamic tides in close binaries and highlight some recent advances
in the comparison between theory and observations.

\section{Basic Assumptions}

We consider a close binary system of stars orbiting around each other
under the influence of their mutual gravitational force. The first
star, with mass $M_1$ and radius $R_1$, is assumed to rotate uniformly
around an axis perpendicular to the orbital plane with a rotational
angular velocity $\vec{\Omega}$ which is small enough to render the
effects of the Coriolis force and the centrifugal force
negligible. The second star, with mass $M_2$, is treated as a point
mass.

In what follows, we denote by $a$ the semi-major axis, by $P_{\rm
orb}$ the orbital period, by $n=2\,\pi/P_{\rm orb}$ the mean motion,
and by $e$ the orbital eccentricity.  The tidal force experienced by
the star is derived from the tide-generating potential
$\varepsilon_T\,W(\vec{r},t)$ which we expand in Fourier series as
\begin{eqnarray}
\lefteqn{ \varepsilon_T\, W \left( \vec{r},t \right) = -
  \varepsilon_T\, {{G\, M_1} \over R_1}\,
  \sum_{\ell=2}^4 \sum_{m=-\ell}^\ell \sum_{k=-\infty}^\infty
  c_{\ell,m,k}\, \left( {r \over R_1} \right)^\ell
  } \nonumber \\
& & \times\,
  Y_\ell^m (\theta,\phi)\, \exp \left[ {\rm i}
  \left( \sigma_T\, t - k\, n\, \tau \right) \right]
  \hspace{3.5cm} \label{pot}
\end{eqnarray}
(Polfliet \& Smeyers 1990; Smeyers, Willems, \& van Hoolst 1998).
Here, $\varepsilon_T = (R_1/a)^3 (M_2/M_1)$ is a small dimensionless
parameter corresponding to the ratio of the tidal force to the gravity
at the star's equator, $G$ is the Newtonian constant of gravitation,
$\tau$ is a time of periastron passage, the factors $c_{\ell,m,k}$ are
Fourier coefficients, the $Y_\ell^m (\theta,\phi)$ are unnormalized
spherical harmonics with respect to a system of spherical coordinates
that is co-rotating with the star, and the $\sigma_T = k\, n + m\,
\Omega$ are forcing angular frequencies with respect to the
co-rotating frame of reference. The Fourier coefficients
$c_{\ell,m,k}$ depend on the orbital eccentricity and are proportional
to $(R_1/a)^{\ell-2}$. They obey the property of symmetry
$c_{\ell,-m,-k} = c_{\ell,m,k}$ and are equal to zero for odd values
of $\ell+|m|$. The coefficients $c_{\ell,m,0}$ are furthermore equal
to zero for $|m| > \ell-1$.

Due to the dependency of the Fourier coefficients $c_{\ell,m,k}$ on
the ratio $R_1/a$, the terms associated with the second-degree
spherical harmonics are dominant. For the remainder of the paper, we
therefore restrict ourselves to the tides generated by the $\ell=2$
terms. The only non-zero Fourier coefficients in the expansion of the
tide-generating potential are then the coefficients $c_{2,-2,k}$,
$c_{2,0,k}$, and $c_{2,2,k}$. These coefficients are independent of
the ratio $R_1/a$ and are solely determined by the orbital
eccentricity $e$.

For a given orbital eccentricity, the coefficients $c_{2,m,k}(e)$
generally decrease with increasing values of $k$. The decrease is
slower for higher orbital eccentricities so that the number of
non-trivially contributing terms in the expansion of the
tide-generating potential increases with increasing values of $e$. For
a given orbital eccentricity, the coefficients $c_{2,-2,k}(e)$ are
furthermore largest at values of $k$ close to those for which $k\,n
\approx \Omega_{\rm P}$, where $\Omega_{\rm P}$ is the orbital angular
velocity at the periastron of the stars' relative orbit. For a more
precise definition and in-depth discussion of the Fourier coefficients
$c_{2,m,k}(e)$, we refer to Smeyers et al. (1998) and Willems (2003).

\section{Resonant dynamic tides}

From Eq.~(\ref{pot}), it follows that for each spherical harmonic
$Y_2^m (\theta,\phi)$, the tide-generating potential generates an
infinite number of partial dynamic tides with forcing angular
frequencies $\sigma_T \ne 0$.  When one of these forcing frequencies
is close to the eigenfrequency of a free oscillation mode, the tidal
action exerted by the companion is enhanced and the oscillation mode
involved is resonantly excited. The possibility of resonances between
partial dynamic tides and free oscillation modes was first suggested
by Cowling (1941) and is particularly relevant for the excitation of
free oscillation modes $g^+$ since their eigenfrequencies are most
likely to be in the range of the forcing frequencies induced in
components of close binaries (Willems 2003).

Neglecting any nonadiabatic effects, the displacement field resulting
from a second-degree resonant dynamic tide associated with the
azimuthal number $m$ and the $k$-th harmonic in the Fourier
decomposition of the tide-generating potential, at the lowest order of
approximation in the small parameter $\varepsilon_T$, is given by
\begin{equation}
\vec{\xi}_T \left( \vec{r},t \right) =  {1 \over \varepsilon}\, 
  {\varepsilon_T \over 2}\, c_{2,m,k}\, {\cal Q}_{2,N}\,
  \vec{\xi}_{2,N}\left( \vec{r}\, \right) \exp \left[ {\rm i}
  \left( \sigma_T\, t - k\, n\, \tau \right) \right] \label{xit}
\end{equation}
(Smeyers et al. 1998). Here, $\varepsilon =
(\sigma_{2,N} - \sigma_T)/\sigma_{2,N}$ is the relative difference
between the forcing frequency of the resonant dynamic tide and the
eigenfrequency of the oscillation mode involved in the resonance,
${\cal Q}_{2,N}$ is the overlap integral, and $\vec{\xi}_{2,N}$ is the
vector of the Lagrangian displacement of the oscillation mode involved
in the resonance. The subscript $N$ denotes the radial order of the
mode. A generalization of Eq.~(\ref{xit}) accounting for the effects
of energy exchange between the mass elements and their surroundings
may be found in Willems, van Hoolst, \& Smeyers (2003).

It follows that even though the time-dependency of the displacement
field is governed by the forcing frequency imposed by the resonant
dynamic tide, the basic properties and general behavior of
$\vec{\xi}_T \left( \vec{r}\, \right)$ are determined by the
eigenfunction of the oscillation mode involved in the resonance. The
amplitude of the displacement field is furthermore determined by the
inverse of the relative frequency difference $\varepsilon$ and by the
overlap integral ${\cal Q}_{2,N}$ which is proportional to the 
work done by the tidal force through the excited 
mode (for a precise definition see,
for example, Zahn 1970, Press \& Teukolsky 1977, Smeyers et
al. 1998). For main-sequence stars, the behaviour of the
overlap-integral is such that the coupling between the oscillation
mode and the tidal force generally weakens with increasing radial
order of the mode.

The effects of resonant dynamic tides on the appearance and evolution
of a binary may be quite substantial. Firstly, the enhanced amplitude
of the tidal motions and the associated radial-velocity variations
provide the most favourable circumstances for the detection of
tidally induced oscillations in close binaries. Willems \& Aerts
(2002) have shown that besides the increase in amplitude, the
radial-velocity variations associated with a single resonance also
become more and more sinusoidal when the forcing frequency becomes
closer to the eigenfrequency of the mode involved in the
resonance. Secondly, the enhanced tidal action can significantly
accelerate the secular evolution of the binary's semi-major axis and
the star's rotational angular velocity (Savonije \& Papaloizou 1983,
1984; Willems et al. 2003). Since the forcing frequencies induced in
the star depend on $n$ and $\Omega$, one would therefore expect
resonances to be fairly short-lived as the binary evolves rapidly
through and away from them. Witte \& Savonije (1999, 2001), however,
have shown that when stellar and orbital evolution are taken into
account simultaneously, the changes in the forcing frequencies and the
eigenfrequencies may compensate one another and the binary may become
locked in a resonance for a prolonged period of time.

Despite the possibility of resonance lockings, firm evidence for the
presence of resonantly excited oscillation modes in components of
close binaries is still scarce. In recent years, however, systematic
large-scale surveys of pulsating stars in close binaries containing
early-type stars have yielded some very promising candidates in the
search for tidally induced oscillations. The stars HD\,177863,
HD\,209295, and HD\,77581, for example, are all multi-periodic
oscillators with observed frequencies equal to an integer multiple of
the orbital frequency\footnote{Note that in the observer's
(non-rotating) frame of reference the condition for a mode with an
observed frequency $\sigma_{\rm obs}$ to be resonantly excited is
$\sigma_{\rm obs} \simeq k\,n$.}. The binary properties and possibly
resonant frequencies of these three systems are summarized in
Table~\ref{stars}.  HD\,209295 is currently believed to have either a
neutron star or white dwarf companion (Handler et al. 2002), while the
companion to HD\,77581 is the well-known Vela X-1 pulsar (Quaintrell
et al. 2003). The nature of the companion to HD\,177863 is currently
unknown.

\begin{table}
\caption{Examples of stars oscillating with observed frequencies
  $\sigma_{\rm obs}$ equal to integer multiples of the orbital
  frequency $n$.}
\label{stars} 
\smallskip
\begin{center}
\begin{tabular}{lcccccc}
\tableline
\noalign{\smallskip}
Name & $M_1$ & $M_2$ & $P_{\rm orb}$ & $e$ & $\sigma_{\rm obs}/n$ &
Refs. \\
\noalign{\smallskip}
\tableline
\noalign{\smallskip}
 HD\,177863 & $3.5M_\odot$ & $1.0M_\odot$ & 11.9\,d  & 0.60 &
 10 & 1, 2, 3 \\
 HD\,209295 & $1.8M_\odot$ & $1.5M_\odot$ &  3.11\,d & 0.35 &
 3, 5, 7, 8, 9 & 4 \\ 
 HD\,77581  & $23\!-\!29M_\odot$ & $1.8\!-\!2.4M_\odot$ & 8.96\,d & 0.09 &
 1, 4 & 5 \\
\noalign{\smallskip}
\tableline
\end{tabular}
\end{center}
{\small References: (1) De Cat et al. 2000; (2) De Cat 2001,
  (3) Willems \& Aerts 2002; (4) Handler et al. 2002; (5) Quaintrell
  et al. 2003.}
\end{table}

\section{Apsidal motion}

In close binaries, the tidal distortion of a star by its companion
gives rise to a perturbation of the external gravitational field which
in turn causes a secular motion of the line of the apses. The most
commonly used formula for the rate of secular apsidal motion was
derived first by Cowling (1938) and subsequently by Sterne (1939)
under the assumption that the orbital period is long in comparison to
the periods of the free oscillation modes of the component stars. The
formula is given by
\begin{equation}
\left( {{d\varpi} \over {dt}} \right)_{\rm clas} = 
  \left( {R_1 \over a} \right)^5 {M_2 \over M_1}\, 
  {{2\,\pi} \over P_{\rm orb}}\, k_2\, 15\, 
  f\!\left( e^2 \right), \label{sterne}
\end{equation}
where $\varpi$ is the longitude of the periastron, $k_2$ is the
apsidal-motion constant, and 
\begin{equation}
f\!\left(e^2\right) = \left({1-e^2}\right)^{-5}\,
  \left({1 + {3\over 2}\, e^2 + {1\over 8}\, e^4 }\right). 
\end{equation}
The apsidal-motion constant $k_2$ depends on the internal structure of
the star and measures the extent to which mass is concentrated towards
the stellar center. The constant takes the value $k_2=0$ in the case
of a point mass and the value $k_2=0.75$ in the case of the
equilibrium sphere with uniform mass density. For main-sequence stars,
$k_2$ is typically of the order of $10^{-3}-10^{-2}$.

Within the framework of the theory of dynamic tides, the rate of
secular apsidal motion is determined by adding the contributions to
the apsidal-motion rate stemming from the various partial static and
dynamic tides generated by the individual terms in the expansion of
the tide-generating potential. The resulting rate can be cast in a
form similar to Eq.~(\ref{sterne}) as
\begin{equation}
\left( {{d\varpi} \over {dt}} \right)_{\rm dyn} = 
  \left( {R_1 \over a} \right)^5 {M_2 \over M_1}\, 
  {{2\,\pi} \over P_{\rm orb}}\, k_{2,{\rm dyn}} 
  \left( \Omega, n, e \right) 15\,f\!\left( e^2 \right)
\end{equation}
(Claret \& Willems 2002). The generalized
apsidal-motion constant $k_{2,{\rm dyn}}$ now renders the response of
the star to the forcing frequencies induced by the tide-generating
potential and therefore depends on the rotational angular velocity
$\Omega$, the mean motion $n$, and the orbital eccentricity $e$. The
constant is determined by numerical integration of the system of
differential equations governing forced nonradial oscillations of a
spherically symmetric equilibrium star. 

Unlike the classical apsidal-motion constant $k_2$, the generalized
apsidal-motion constant $k_{2,{\rm dyn}}$ can be negative as well as
positive so that periastron recessions can occur as well as periastron
advances. Periastron recessions typically arise for close resonances
when the forcing frequency of the resonant dynamic tide is larger than
the eigenfrequency of the oscillation mode involved in the
resonance. When the forcing frequency of the resonant dynamic tide is
smaller than the eigenfrequency of the oscillation mode involved in
the resonance, the apsidal motion is generally a periastron advance.

In the limiting case of long orbital and rotational periods, the rates
of secular apsidal motion derived by Cowling (1938) and Sterne (1939)
agree with the formula established within the framework of the theory
of dynamic tides up to high orbital eccentricities (Quataert, Kumar,
\& Ao 1996; Smeyers et al. 1998). The agreement rests on the property
that when all forcing frequencies $\sigma_T$ tend to zero, the
generalized apsidal-motion constant $k_{2,{\rm dyn}}$ tends to the
classical apsidal-motion constant $k_2$.  For binaries with shorter
orbital and/or rotational periods, deviations arise due to the
increasing role of stellar compressibility at higher forcing
frequencies and due to resonances of dynamic tides with free
oscillation modes (Smeyers \& Willems 2001).

\begin{figure}
\plottwo{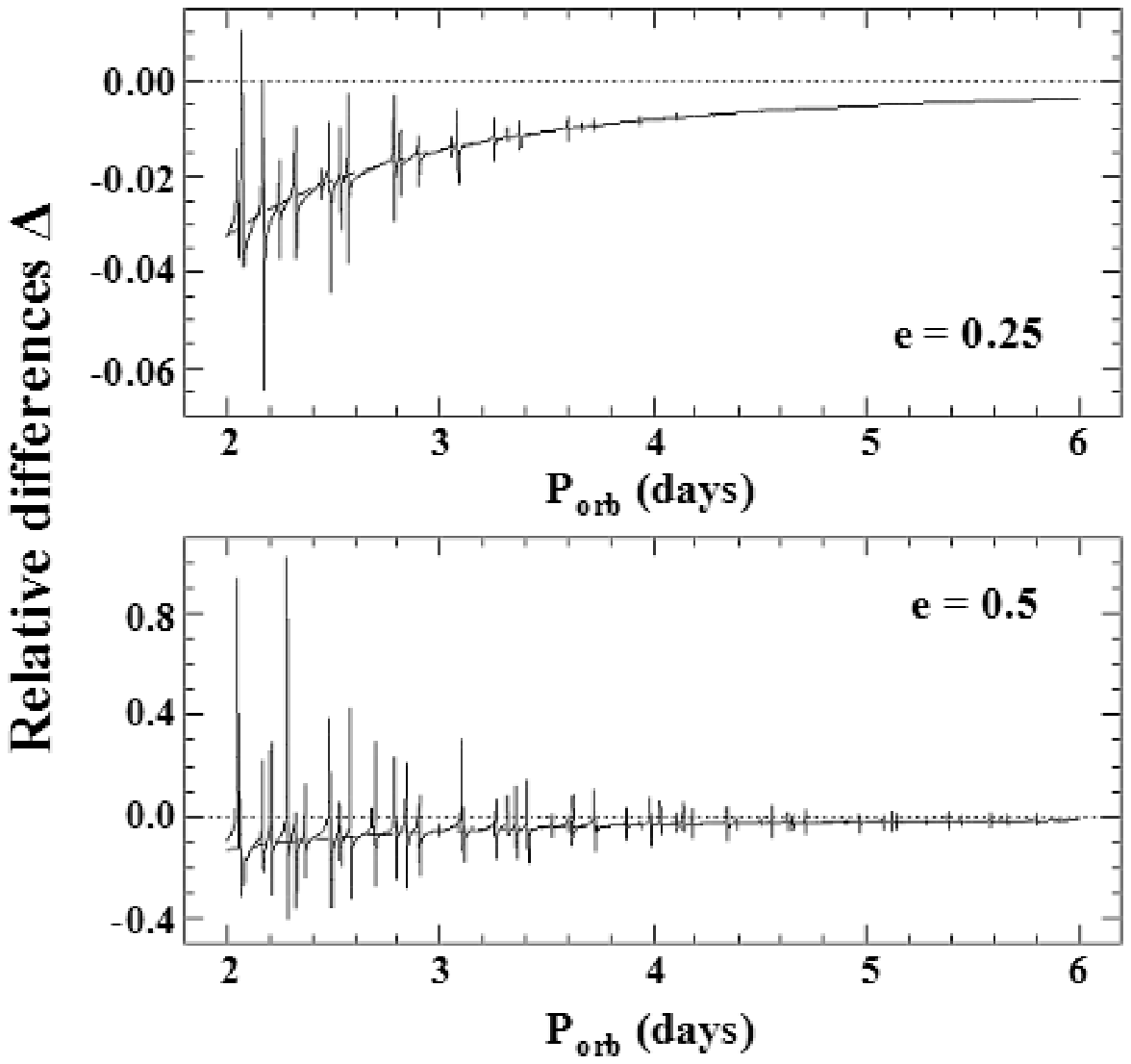}{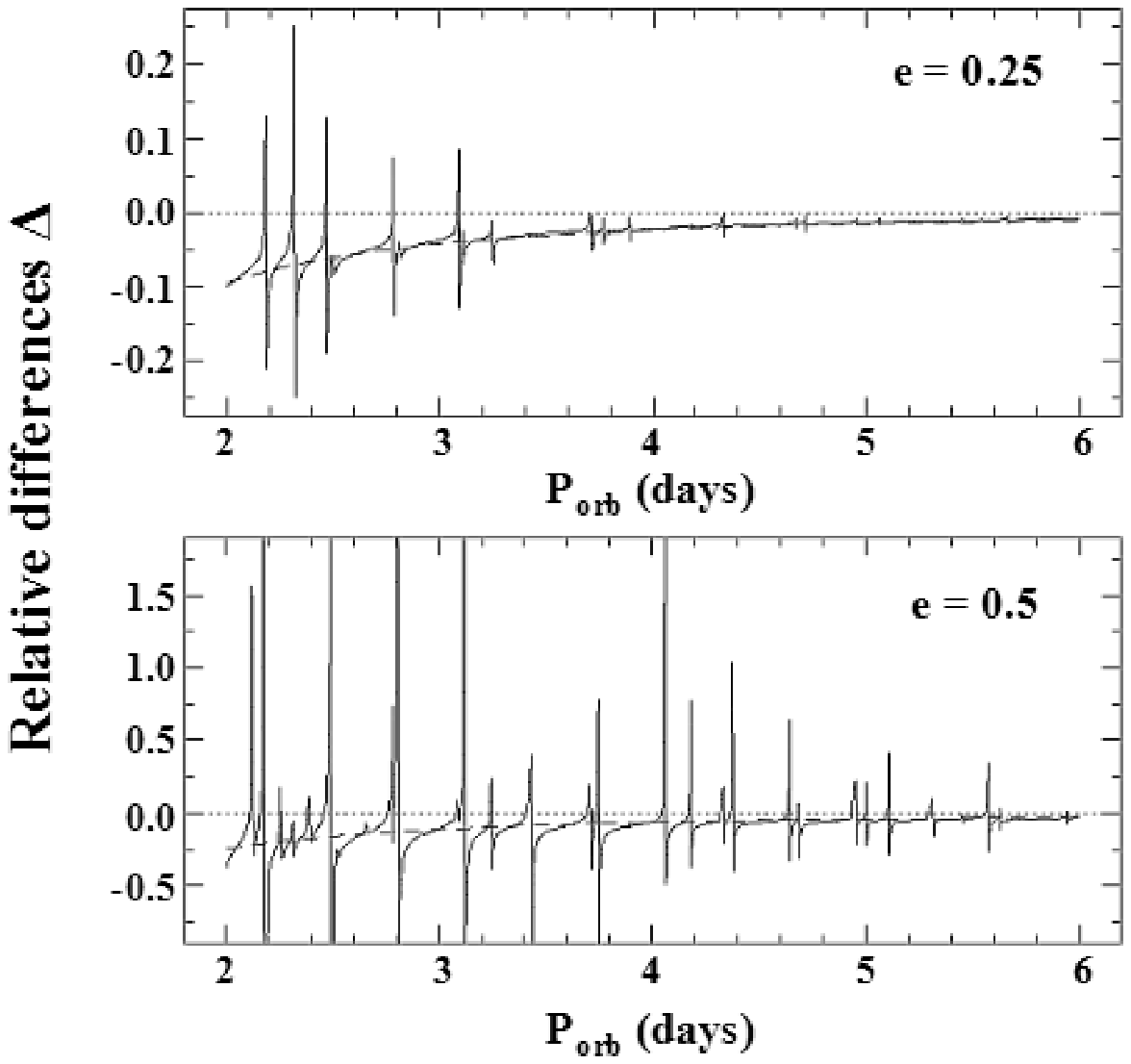}
\caption{Variations of the relative difference $\Delta$ as a function
  of the orbital period for a $5\,M_\odot$ (left-hand panel) and a 
  $20\,M_\odot$ (right-hand panel) ZAMS star, and for the orbital
  eccentricities $e=0.25$ and $e=0.5$. The rotational angular velocity
  was set to $\Omega = 0.01\,n$ so that the forcing frequencies are
  predominantly determined by the mean motion $n$. The response of the
  star to the forcing frequencies was determined in the isentropic
  approximation. The amplitudes near resonances are artificially
  limited to a finite value as explained in Smeyers \& Willems
  (2001).}  
\label{aps1}
\end{figure}

A useful quantity to illustrate the deviations between the
apsidal-motion rate derived by Cowling (1938) and Sterne (1939) and
the rate derived within the framework of the theory of dynamic tides
is the relative difference
\begin{equation}
\Delta = {{(d\varpi/dt)_{\rm clas} 
   - (d\varpi/dt)_{\rm dyn}} \over 
   {(d\varpi/dt)_{\rm dyn}}}
   = {{k_2 - k_{2,{\rm dyn}}} \over {k_{2,{\rm dyn}}}}.
\end{equation}
The variations of $\Delta$ as a function of $P_{\rm orb}$ are
displayed in Fig.~\ref{aps1} for a $5\,M_\odot$ and a $20\,M_\odot$
ZAMS star, and for $\Omega = 0.01\,n$. The relative differences are
mostly negative so that for binaries with shorter orbital periods the
classical formula yields somewhat too small values for the rate of
secular apsidal motion, and thus somewhat too long apsidal-motion
periods. The peaks observed at shorter orbital periods are caused by
resonances of dynamic tides with free oscillation modes $g^+$ of the
tidally distorted star. These peaks are superposed on a basic curve
which represents the systematic deviations caused by the increasing
role of the stellar compressibility at shorter orbital periods.

The systematic deviations due to the stellar compressibility increase
with increasing mass of the ZAMS star. This behaviour is related to
the decrease in the amount of central condensation and the associated
increase in the impact of tides with increasing stellar mass. The
systematic deviations also become larger with increasing orbital
eccentricity due to the larger number of higher-frequency tides
contributing to the tide-generating potential.  The number of resonant
dynamic tides on the other hand tends to decrease with increasing
stellar mass. The reason for this is that ZAMS stars with higher
masses have smaller radiative envelopes which causes their
eigenfrequencies to be larger and more widely spaced than those of a
lower-mass ZAMS star. For a more extended discussion of the relative
differences $\Delta$ for ZAMS stars, we refer to Smeyers \& Willems
(2001).

The extent of the deviations caused by the compressibility of the
stellar fluid also depends on the evolutionary stage of the star
(Willems \& Claret 2002). The dependency is primarily through the
evolution of the radius and the associated change of the star's
dynamical time scale. In particular, as the star evolves on the main
sequence, the radius and the dynamical time scale increase so that in
relative terms the orbital period becomes shorter in comparison to the
star's dynamical time scale. Correspondingly, the forcing frequencies
become larger when expressed in units of the inverse of the star's
dynamical time scale.  The deviations due to the compressibility of
the stellar fluid are therefore larger for a model with a larger
radius.  The evolution of the star furthermore also affects the
deviations caused by the resonances of dynamic tides with free
oscillation modes of the component stars. In particular, the effects
of the resonances tend to be larger for stars near the end of
core-hydrogen burning than for stars on the zero-age main
sequence. This effect is again related to the size of the radiative
envelope, which increases which increasing age of the star on the main
sequence.

The dependency of the apsidal-motion rate on the stellar structure
makes the comparison between theoretically predicted and
observationally derived ap\-si\-dal-motion rates an excellent tool to
probe the interior of close binary components (see, e.g., Claret \&
Gim\'enez 1993).  In order to reliably compare theory and
observations, the contributions to the apsidal-motion rate due to
stellar rotation and general relativity must be taken into account in
addition to the contribution due to tides.

Stellar rotation contributes to the apsidal motion in two
ways. Firstly, the distortion of the star due to rotation contributes
to the perturbation of the external gravitation field. The associated
rate of secular apsidal motion again depends on the star's internal
structure through the apsidal-motion constant $k_2$ and on the stellar
and orbital dimensions through the factor $(R_1/a)^5$. The rate is
given by
\begin{equation}
\left( {{d\varpi} \over {dt}} \right)_{\rm rot} = 
  \left( {R_1 \over a} \right)^5 {{M_1 + M_2} \over M_1}\, 
  {{2\,\pi} \over P_{\rm orb}}\, 
  \left( {\Omega \over n} \right)^2 k_2\, 
  \left( 1-e^2 \right)^{-2}  \label{rot1}
\end{equation}
(e.g. Sterne 1939). Secondly, rotation tends to increase a star's central
condensation and therefore decreases the apsidal-motion constant
$k_2$. Claret (1999) showed that the difference in $\log k_2$ can be
approximated as
\begin{equation}
\Delta \log k_2 = -0.87 \left( {{2\, R_1\, \Omega^2} \over 
  {3\, G\, M_1 / R_1^2}} \right) + 0.0004.  \label{rot2}
\end{equation}

The general relativistic contribution to the rate of secular apsidal
motion on the other hand is independent of the star's radius and
internal structure. Instead, it depends on the total system mass
$M_1+M_2$, the orbital semi-major axis $a$, and the orbital
eccentricity $e$: 
\begin{equation}
\left( {{d\varpi} \over {dt}} \right)_{\rm GR} = {{3\,G} \over c^2}\, 
  {{2\,\pi} \over P_{\rm orb}}\, {{M_1+M_2} \over 
  {a \left( 1-e^2 \right)}} \label{gr}
\end{equation}
(Levi-Civita 1937, Peters 1964). The general relativistic
apsidal-motion rate is often small, but may become dominant in
binaries with high orbital eccentricities (e.g. V1143\,Cyg with
$M_1=1.4\,M_\odot$, $M_2=1.3\,M_\odot$, $P_{\rm orb}=7.6$\,days, and
$e=0.54$). 

Testing theories of stellar structure and evolution by comparing
predicted and observed apsidal-motion rates furthermore requires both
accurate observations and high-quality stellar models (Claret \&
Willems 2002). Due to the strong dependence of the apsidal-motion rate
on the ratio $R_1/a$, the observations must yield accurate
determinations of the stellar and orbital dimensions, while the
stellar models must be able to simultaneously match the
observationally inferred masses, radii, and effective temperatures of
both stars at the same age. In the context of the theory of dynamic
tides, the orbital and rotational periods must furthermore be known
with a high degree of accuracy in order to precisely determine the
forcing frequencies induced in the binary components. While the
accurate determination of the orbital period usually poses no problem,
the determination of the rotational period of close binary components
often remains a challenge.

\begin{figure}
\plottwo{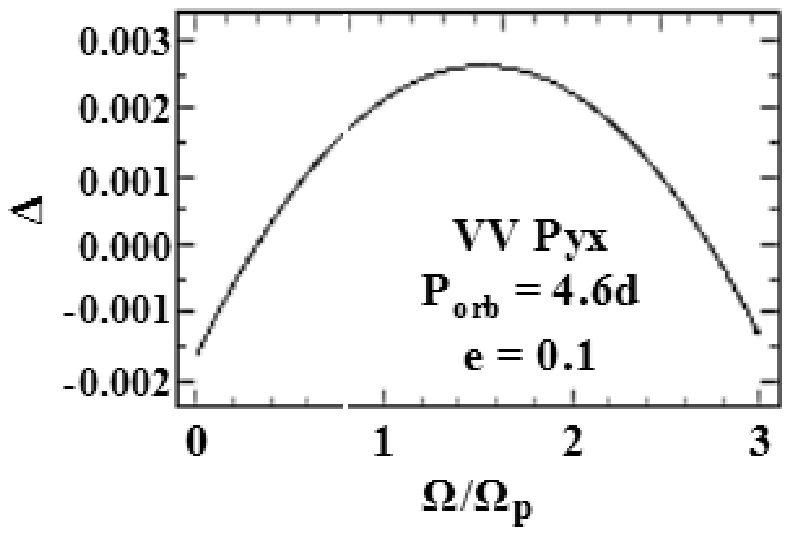}{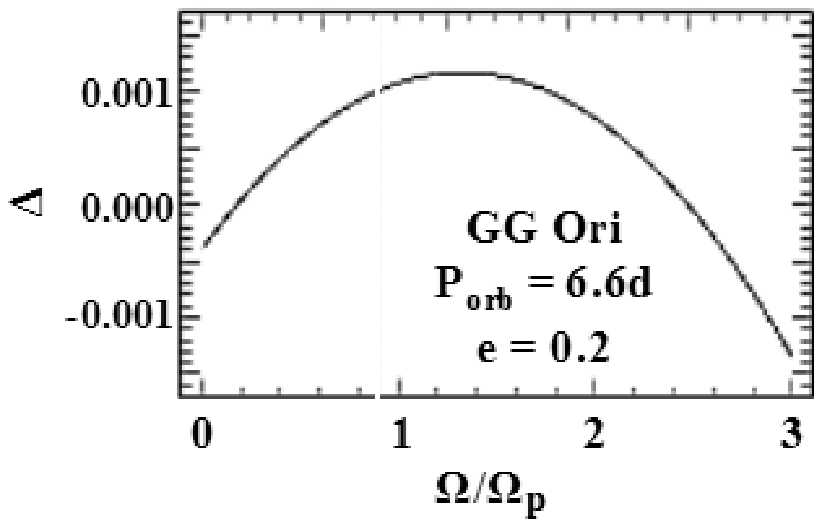}
\caption{Variations of the relative difference $\Delta$ as a function
  of the ratio of the rotational angular velocity to the orbital
  angular velocity at periastron for the binaries VV\,Pyx and
  GG\,Ori.}
\label{vv}
\end{figure}

\begin{figure}
\plottwo{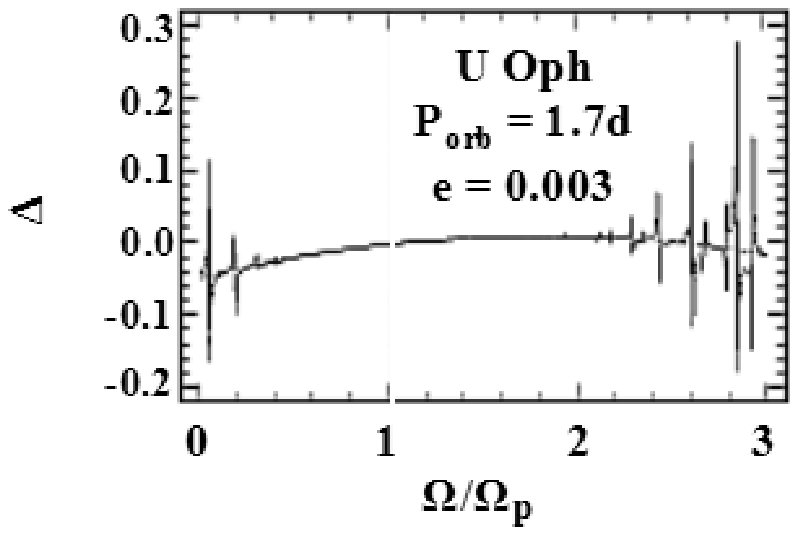}{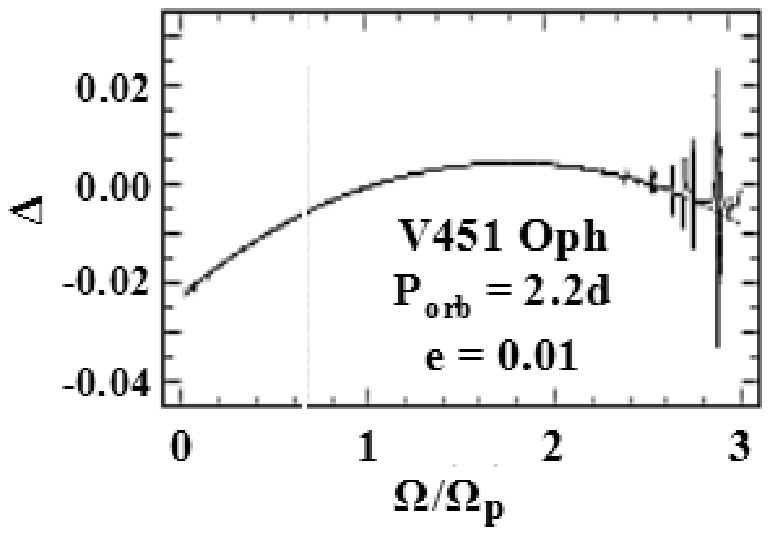}
\caption{Variations of the relative difference $\Delta$ as a function
  of the ratio of the rotational angular velocity to the orbital
  angular velocity at periastron for the binaries U\,Oph and
  V451\,Oph.}
\label{uo}
\end{figure}

The lack of solid rotational angular velocity determinations may yield
large uncertainties in the calculation of theoretical apsidal-motion
rates within the framework of the theory of dynamic tides. This is
illustrated in Figs.~\ref{vv} and ~\ref{uo} where the relative
differences $\Delta$ for the eclipsing binaries VV\,Pyx, GG\,Ori,
U\,Oph, and V451\,Oph are displayed as a function of the rotational
angular velocity $\Omega$ in units of the orbital angular velocity
$\Omega_{\rm P}$ at periastron. In the cases of VV\,Pyx and GG\,Ori,
the relative differences are small and entirely due to the effects of
the stellar compressibility for the entire range of rotational angular
velocities considered. In the cases of U\,Oph and V451\,Oph, on the
other hand, the answer to the question of whether or not dynamic tides
play a role in the determination of the contribution of tides to the
apsidal-motion rate strongly depends on the value of the star's
rotational angular velocity. It can furthermore be seen that a small
error in $\Omega$ can mean the difference between being in a resonance
or not. 

Claret \& Willems (2002) showed that when all the aforementioned
uncertainties and contributions to the apsidal-motion rate are taken
into account, a generally good agreement between theoretically derived
and observationally inferred apsidal-motion rates is found. The
authors furthermore found no systematic deviations between predicted
and observed apsidal-motion rates as a function of the evolutionary
stage of the stars. Despite the good agreement, one should keep in
mind though that the theoretical treatment of the problem is far from
complete. The detailed effects of rotation on the tidal motions and
the internal structure of the stars, for example, has yet to be
consistently included in the calculations. Claret \& Willems (2003)
performed a first preliminary exploration of the effects of the
modified internal structure and found that huge differences may arise
from using rotating stellar models instead of non-rotating ones. Their
exploration, however, did not take into account the effects of the
Coriolis force, which form an indispensable ingredient to the problem
but also significantly complicate the theoretical treatment (see Rocca
1987, 1989). 

\section{Conclusions}

In recent years, significant progress has been made in the theory of
dynamic tides in close binaries as well as in the observations of
tidally active close binaries. From the theoretical side, the effects
of dynamic tides have been incorporated into the classical
apsidal-motion test to stellar structure and evolution, while from the
oservational side more and more evidence is arising supporting the
theoretically predicted resonant excitation of free oscillation
modes. As more and more data becomes available from large-scale
surveys of pulsating stars in close binaries (e.g. Harmanec et
al. 1997; Aerts et al. 1998, 2000), the full potential of tides as a
tool to test theories of stellar and binary evolution may soon be
within our reach. The known sample of eclipsing binaries showing
apsidal-motion may furthermore significantly benefit from the large
number of new systems expected to be found from large-scale surveys
looking for exoplanets transiting the disk of their host star (see Horne
2002 and Charbonneau 2003 for an overview of ground-based transit
searches). 

\acknowledgments{BW is grateful for the financial support supplied by
  the organizing committee and NASA ATP grant NAG5-13236 to Vicky
  Kalogera.}

\end{document}